\documentclass[prb,aps,groupedaddress,twocolumn,reprint,showpacs,preprintnumbers,amsmath,amssymb]{revtex4-1}

\usepackage{graphicx, bm, color, hyperref}

\begin{document}

\title{Transient Response of the Cavity-Magnon-Polariton}
\author{Christophe Match}
\author{Michael Harder}
\email{michael.harder@kpu.ca}
\thanks{Current Address: Department of Physics, Kwantlen Polytechnic University, 12666 72 Avenue, Surrey, BC V3W 2M8 Canada.}
\author{Lihui Bai}
\thanks{Current Address: Shandong University, 27 Shandanan Road, Jinan, 250100 China.}
\author{Paul Hyde}
\author{Can-Ming Hu}

\affiliation{Department of Physics and Astronomy, University of Manitoba, Winnipeg, Canada R3T 2N2}

\date{\today}

\begin{abstract}
The Rabi oscillations of a coupled spin-photon system are experimentally studied using time domain microwave transmission measurements.  An external magnetic bias field is applied to control the amplitude and frequency of Rabi oscillations, which are a measure of the coherent spin-photon energy exchange and coupling strength respectively.  The amplitude and frequency control is described in the time domain as the transient response of a two-level spin-photon system.  This approach reveals the link between steady state and transient behaviour of the cavity-magnon-polariton, which is necessary to interpret time domain cavity-spintronic experiments.  This transient model also allows us to understand and quantify the polariton mode composition, which can be controlled by the bias field. 
\end{abstract}

\maketitle

\section{Introduction}
Motivated by the potential for novel quantum information and spintronic technologies, much work has recently been devoted to the study of exchange coupled spin ensembles which display long coherence times and can easily be manipulated by microwave magnetic fields. \cite{Harder2018, Huebl2013a, Tabuchi2014a, Zhang2014, Soykal2010,  Bai2015, Maier-Flaig2016, Bai2017, Cao2014, Yao2015,  Harder2016b, Zhang2015e, Haigh2015a, Bourhill2015a}  In the frequency domain both optical \cite{Huebl2013a, Tabuchi2014a, Zhang2014} and electrical detection experiments \cite{Bai2015, Maier-Flaig2016, Bai2017} as well as various theoretical approaches \cite{Soykal2010, Zhang2014, Bai2015, Cao2014, Yao2015, Harder2016b} have elucidated the key properties of such systems, such as a dispersion anti-crossing and line width evolution, which result when strong electrodynamic interactions lead to the formation of a coupled spin-photon system -- the cavity-magnon-polariton (CMP).  

The interest in cavity-magnon-polaritons is tied to their underlying coherence properties which hybridize the spin and photon states, resulting in an excellent transducer.  For example, the realization of a gradient memory utilizing magnon dark modes, \cite{Zhang2015g} demonstration of magnon-qubit coupling, \cite{Tabuchi2015b} and the development of optical to rf conversion techniques \cite{Haigh2015b, Zhang2015b, Hisatomi2016} have been based on CMP coherence.  Of course the coherent manipulation of quantum states is also a key requirement for quantum computation.  In this regard important progress has been made in the strong coupling of solid state atom-cavity systems, such as quantum dots and nitrogen-vacancy (NV) centres. \cite{Kloeffel2013, Barclay2014}  For example single-spin-photon cooperativities of $C\sim 15$ have been achieved in silicon quantum dots, \cite{Mi2017, Samkharadze2018} and $C \sim 27$ can be reached in ensembles of diamond NV centres. \cite{Kubo2010}  A key to achieving such high cooperatives is the use of spin ensembles.  In this regard magnon systems have great potential due to their collective nature, resulting in controllable cooperativities routinely in excess of $C \sim 1000$ \cite{Harder2018} and even reaching the ultra strong regime with $C > 10^4$. \cite{Zhang2014}

On the other hand the manifestation of coherence properties is also important for spintronic applications, enabling long distance spin current manipulation. \cite{Bai2017}  Although it is known that time domain information, in particular the systematic study of Rabi oscillations, is important to understand coherence \cite{Stievater2001, Martinis2002}, so far our understanding of the photon-magnon system has come from frequency domain studies.  Beyond coherence, access to time domain information could improve our understanding of spin current decay and modification due to coupling.  However, perhaps the most pressing motivation for a time domain CMP study lies in the potential of THz CMP phenomena.  Although spin-photon coupling has been well studied in ferrimagnetic materials at GHz frequencies, where continuous wave (CW) experiments excel, the first experimental investigation of CMP phenomena in antiferromagnets has just appeared \cite{Bialek2019}.  Unlocking the true potential of the antiferromagnetic CMP will likely require THz time domain studies.\cite{Satoh2010, Baltz2018}

In this work our goal is to elucidate the relationship between CW and time domain CMP dynamics, and to better understand the spin-photon energy exchange.  To address this problem we have used the eigensolution of the spin-photon system to quantitatively describe the transient behaviour of the CMP.  Our model has been experimentally verified using a vector-network-analyzer based time domain measurement, however our model does not rely on such measurements, and is not limited by them.  These measurements demonstrate Rabi oscillation amplitude and frequency control using an external bias field and allow us to extract the energy stored in both the spin and photon systems.  The energy distribution is determined by the initial state of the system which we can control via the bias field and is related to the polariton mode composition.  Therefore by exploiting the spin-photon coherence as revealed through Rabi oscillations, we provide a method to characterize and control the energy distribution between spin and photon states while also characterizing the polariton mode composition.  At the same time we clarify the connection between CW and time domain response, which could benefit future time domain studies of CMP dynamics, in particular the THz studies of antiferromagnets.

This paper is organized as follows: in Sec \ref{sec:exp} we introduce the experimental setup and examine the CMP response in both frequency and time domains.  In Sec \ref{sec:model} we describe our model of the spin-photon transient response.  We then demonstrate the bias field control of both Rabi frequency and amplitude, and more generally the time domain analysis procedure, in Sec. \ref{sec:rabi}.  In Sec. \ref{sec:comp} we outline the method used to extract the polariton mode composition and its relation to the energy exchange.  Finally in Sec. \ref{sec:conclusions} we conclude with a brief summary.

\section{Time Domain Measurements of the Spin-Photon System} \label{sec:exp}

\begin{figure}[!t]
\centering
\includegraphics[width=\linewidth]{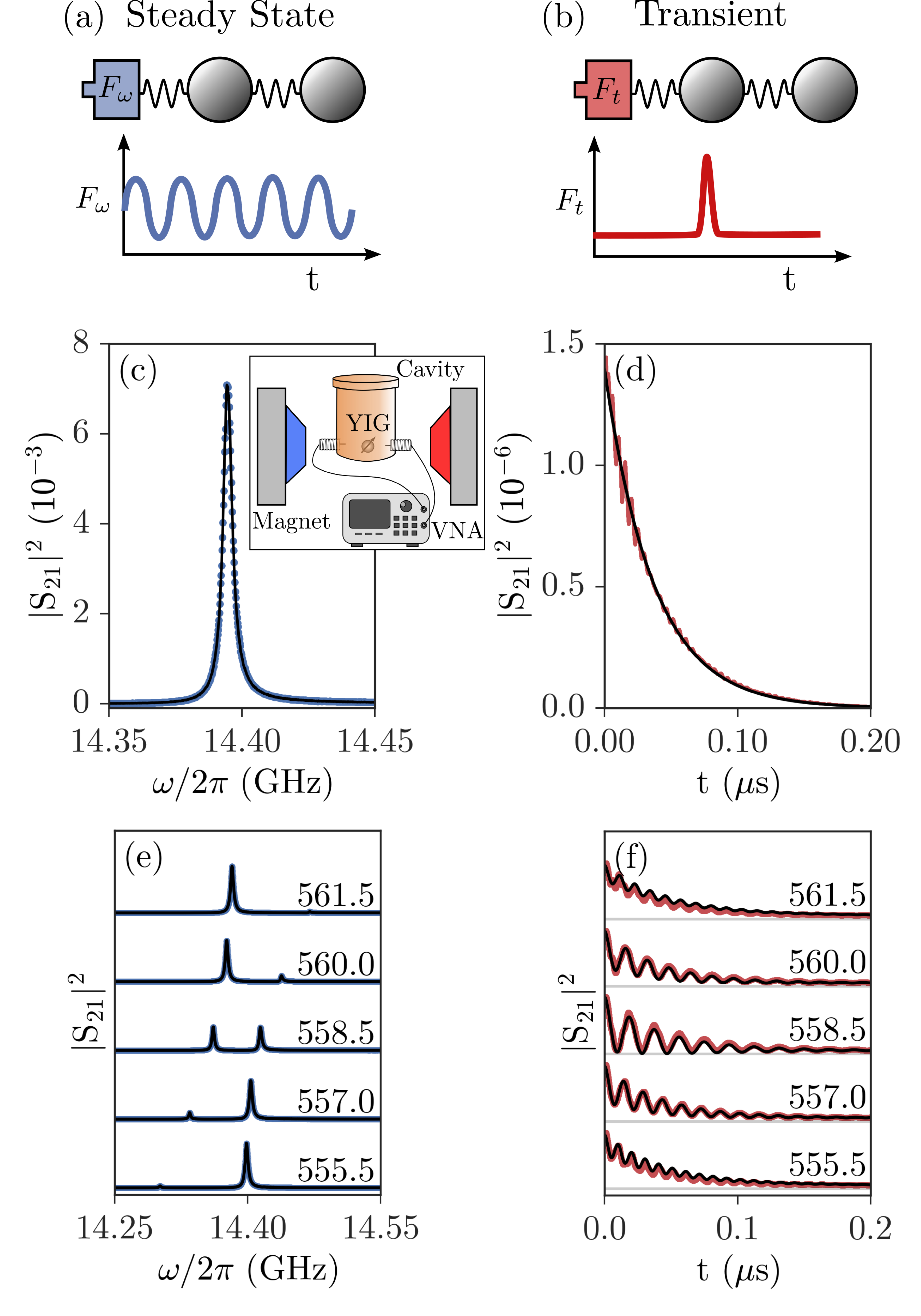}
\caption{Transient and steady state characteristics of the spin-photon system.  (a) Schematically the steady state response is generated by an oscillating driving force while (b) the transient response may result from a pulsed excitation.  Experimentally the time domain response may be obtained using a Fourier transform of the frequency domain VNA data. (c) Transmission spectra of an uncoupled cavity mode as viewed in the frequency and (d) time domains.   The inset of panel (c) shows the experimental setup.  (e) Transmission spectra of the coupled spin-photon system in the frequency and (f) time domains.   In panels (c) and (e) the blue symbols are experimental frequency domain data while the black curves are calculations according to established CMP models. \cite{Harder2016b}  In panels (d) and (f) the red symbols are experimental time domain data, while the solid black curves are fitting results according to Eq. \eqref{eq:h2}.}
\label{fig1}
\end{figure}

To perform both frequency and time domain measurements of the spin-photon system, we placed an yttrium-iron-garnet (YIG) sphere (diameter = 1 mm) at the bottom, outer edge of a cylindrical microwave cavity (diameter = 25.0 mm, height = 40.5 mm) made of oxygen free copper, as shown in the inset of Fig. \ref{fig1} (c).  An external bias field was then applied and the microwave transmission was measured using an Agilent N5230C vector network analyzer (VNA).  When operating in the frequency domain the VNA provides a continuous wave microwave signal which drives the spin-photon system at a constant frequency and probes the response function at the same frequency, as schematically illustrated in Fig. \ref{fig1} (a).  In other words, in the frequency domain microwave transmission measurements probe the steady state response of the spin-photon system.   However the VNA, through in-built Fourier transformation software, can also access the time domain response.  In such a measurement the system is effectively driven by a pulse, as illustrated in Fig. \ref{fig1} (b) and therefore time domain measurements probe the transient response of the spin-photon system.  Although in our experiment we have accessed such information through a Fourier transform, the time domain S-parameter information is equivalent to what one would obtain through time-domain reflectometry (TDR). \cite{TeppatiBook} The pulsed input of TDR or other techniques like pulsed inductive microwave magnetometry (PIMM) allows direct access to dime domain information.  In fact similar approaches have allowed direct storage and retrieval of information in a CMP system. \cite{Zhang2015g}  However importantly the analysis procedure described in Sec. \ref{sec:model} is based on a consideration of CMP eigenmodes and therefore does not rely on the details of the experimental technique.  In this sense the transient response is universal, and our approach would be applicable, for example, to THz time domain studies.

To characterize the spin and photon subsystems we independently measured the empty cavity transmission and ferromagnetic resonance (FMR) dispersion.  Fig. \ref{fig1} (c) shows the microwave transmission of the empty cavity in the frequency domain ($|\text{S}_{21}|^2$) with zero magnetic bias field.  A sharp peak occurs at the TM$_{013}$ resonance frequency $\omega_c/2\pi = 14.391$ GHz.  A Lorentzian fit of the TM$_{013}$ mode, shown as the black curve in Fig. \ref{fig1} (c), yields the half-width-half-maximum, $\Delta \omega/2\pi = 2.14$ MHz, corresponding to a damping coefficient of $\beta = \Delta\omega/\omega_c = 1.49\times10^{-4}$ ($Q = 3400$). The asymmetry which can be observed in panel (c) is due to the presence of an additional low intensity cavity mode at 14.190 GHz.  The amplitude of this mode is $\sim 2\%$ of the 14.391 GHz mode amplitude, and therefore has a negligible influence on the magnon-photon coupling, aside from a slight line shape distortion. \cite{Yao2015}  Measurements of the YIG FMR (not shown) follow the linear relation $\omega_r/2\pi = \gamma \left(H + H_A\right)$, with a gyromagnetic ratio, $\gamma = 28 \times 2 \pi ~\mu_0$ GHz/T and shape anisotropy field $\mu_0H_A = -44$ mT.  Additionally the Gilbert damping is found to be $\alpha = 0.81\times10^{-4}$.  From these measurements we expect that $\omega_c = \omega_r$ at $\mu_0$H$_c$ = 558.5 mT.  We will see later that at this coincidence point, $H_c$, the energy exchange between the spin and photon subsystems is maximized and the time averaged energy is equally distributed.

In contrast to the frequency localized peak of Fig. \ref{fig1} (c), in Fig. \ref{fig1} (d) we show the time domain transmission signal of the empty cavity which follows an exponential decay with time constant $\tau/2 = 37$ ns (black curve).  This decay constant is related to the mode line width, $\Delta \omega/2\pi = \left(2\pi \tau\right)^{-1} = 2.15$ MHz, and is in excellent agreement with the frequency domain measurement.  The high frequency, low amplitude oscillations in Fig. \ref{fig1} (d) are a beating effect which occurs between the dominant cavity mode at $\omega_c/2\pi = 14.391$ GHz and the additional low amplitude cavity mode at $\omega_{c1} = 14.190$ GHz.  The predicted beat frequency of $\left(\omega_c - \omega_{c1}\right)/2\pi = 0.200$ GHz corresponds to the observed oscillation frequency of 0.202 GHz.  Furthermore, since $\Delta \omega_c > \omega_{c1}$ and the low intensity mode amplitude is approximately 2\% of the dominant 14.391 GHz mode, the low intensity oscillation amplitude will decay in time following Eq. \eqref{eq:h2}, as observed.

When the bias field is tuned near the coincidence point energy may be effectively transferred between the spin and photon subsystems and a strong hybridization occurs.  Fig. \ref{fig1} (e) shows the frequency domain transmission measurements near $H_c$.  The two peaks correspond to the two eigenmodes of the CMP, and show an anti-crossing, having equal amplitude at $H_c$.  Their separation at this point is $\omega_\text{gap}/2\pi$ = 53 MHz, from which the spin-photon coupling strength, $\kappa = \omega_\text{gap}/\omega_c = 0.0037$ can be determined. \cite{Harder2016b}  The cooperativity of our spin-photon system is therefore $C = \kappa^2/\alpha\beta \approx 1200$.  As the bias field is detuned from the coincident point the dominant mode becomes more cavity-like.  The solid black curve in Fig. \ref{fig1} (e) is a calculation according to the steady state response of a two-level system. \cite{Harder2016b}  In Fig. \ref{fig1} (f) we show the analogous transmission spectra in the time domain.  Physically the two CMP modes will create two GHz oscillations which differ by the MHz Rabi frequency.  This means that the superposition of both modes should produce a microwave frequency beating.  The VNA experiment we perform measures the envelope function of the microwave beating, as illustrated in Fig. \ref{fig1} (f).  In particular the GHz time dependence of the microwave fields is not measured.  However the MHz Rabi oscillations, signalling the coherent spin-photon coupling, are evident.  Interestingly the envelope function and Rabi oscillations display distinctive behaviour as the bias field is tuned.  First, the oscillation frequency can be controlled, reaching a minimum at $\mu_0H_c = 558.5$ mT.  Second, the extinction rate, signalling the energy transferred from the photon to the spin system, can be controlled, reaching a maximum at $H_c$.  This indicates that the maximal coherent energy transfer occurs at $H_c$ as we will quantify later.  The black curves are fits obtained by modelling the transient response of the two-level system as described below.

\section{Modelling the Transient Response} \label{sec:model}

The coupled spin-photon system can be described by the two-level Hamiltonian \cite{Huebl2013a, Zhang2014, Bai2015, Cao2014, Harder2016b, Harder2017}

\begin{equation}
H = \left(\begin{array}{cc}
\omega_c - i \beta \omega_c & \kappa \omega_c \\
\kappa \omega_c & \omega_r - i \alpha \omega_c \end{array}\right). \label{eq:hamiltonian}
\end{equation}
In previous studies the transient response has been neglected in order to study the steady state solutions which describe the response function in the frequency domain.  However to describe the time domain response and characterize the Rabi oscillations we wish to find the transient solutions.  
The normal modes of Eq. \eqref{eq:hamiltonian} will take the form $X_\pm \propto e^{i \tilde{\omega}_\pm t}$, where the eigenfrequencies are
\begin{equation}
\tilde{\omega}_\pm = \frac{1}{2}\left[\tilde{\omega}_c + \tilde{\omega}_r \pm \sqrt{\left(\tilde{\omega}_r - \tilde{\omega}_c \right)^2 + 4 \kappa^2 \omega_c^2} \right],\label{eq:eigen}
\end{equation}
with $\tilde{\omega_c} = \omega_c - i \beta \omega_c$ and $\tilde{\omega}_r = \omega_r - i \alpha \omega_c$.  These polariton modes can be related to the spin and photon modes, $m$ and $h$ respectively, through a coordinate transformation 

\begin{equation}
\left(\begin{array}{c} 
h \\
m \end{array} \right) = \left(\begin{array}{cc}
C_{11} & C_{12} \\
C_{21} & C_{22} \end{array} \right) \left(\begin{array}{c}
X_+ \\
X_- \end{array}\right). \label{eq:coordtransform}
\end{equation}
We consider an excitation pulse at $t = 0$ (in our case effectively obtained through Fourier transformation) which drives a microwave field in the cavity which in turn drives a spin precession in the ferrimagnetic sample.  Therefore $h|_{t = 0} = A$ and $m|_{t = 0}  = 0$, where $A^2$ is related to the microwave power input to the cavity, determined by the microwave cables, pins and VNA.  Using these initial conditions, in addition to the constraint that the spin and photon modes must satisfy the eigenvalue problem defined by Eq. \eqref{eq:hamiltonian}, the coordinate transformation between the polariton and the spin/photon basis is determined to be,   

\begin{equation}
\left(\begin{array}{cc}
C_{11} & C_{12} \\
C_{21} & C_{22} \end{array} \right) = 
A \left(\begin{array}{cc}
1-c & c \\
\sqrt{c-c^2} & -\sqrt{c-c^2} \end{array} \right),
\label{eq:coordtransformc}
\end{equation}
where
\begin{equation}
c = c\left(H\right) = \frac{1}{2} \left[1-\frac{\tilde{\omega}_c - \tilde{\omega}_r}{\sqrt{\left(\tilde{\omega}_c - \tilde{\omega}_r\right)^2 + 4 \kappa^2 \omega_c^2}}\right]. \label{eq:c}
\end{equation}
Since the imaginary component of $c$ is small, the transient response of the microwave transmission can be fully determined by taking the coordinate transformation $C_{ij}$ real,
\begin{align}
|\text{S}_{21}|^2 \propto |h|^2 &= C_{11}^2 e^{-2 t/\tau_+} + 2 C_{11} C_{12} e^{-t/\tau_+} e^{-t/\tau_-} \cos\left(\Gamma t\right) \nonumber\\
& ~~~+ C_{12}^2 e^{-2t/\tau_-}. \label{eq:h2}
\end{align}
We note that keeping the complex components of $C_{ij}$ would add an additional low amplitude $\sin\left(\Gamma t\right)$ term to Eq. \eqref{eq:h2}.  Analogously for $m$ we find,
\begin{equation}
|m|^2 = C_{21}^2 \left[e^{-2 t/\tau_+} - 2 e^{-t/\tau_+} e^{t/\tau_-} \cos\left(\Gamma t\right) + e^{-2t/\tau_-}\right]. \label{eq:m2}
\end{equation}
Note that the equivalence between $\text{S}_{21}$ and $h$ is due to the fact that we have defined the microwave field to include a proportionality constant $A$ which characterizes the impedance matching between the microwave cavity and VNA. \cite{Harder2016b} Experimentally $A$ is treated as a fitting parameter as described below.  Here $\Gamma$ is the generalized Rabi frequency

\begin{equation}
\Gamma = \sqrt{\left(\tilde{\omega}_r - \tilde{\omega}_c\right)^2 + \Omega^2}, \label{eq:rabi}
\end{equation}
which depends on the frequency detuning $\delta = \left(\tilde{\omega}_r - \tilde{\omega}_c\right)$ and the Rabi frequency $\Omega = 2 \kappa \omega_c$, and  $\tau_\pm$ are the decay rates of the two polariton branches, which depend on the magnetic bias field and are related to the mode linewidths according to $\Delta\omega_\pm = \tau_\pm^{-1}$ where $\Delta \omega_\pm = \text{Im}\left(\tilde{\omega}_\pm\right)$ is the half-width at half-maximum for the respective mode.  Calculations according to Eq. \eqref{eq:h2} are plotted as solid black curves in Fig. \ref{fig1} (d).  While this generalized Rabi frequency is dependent on the bias field, $H$, the coupling strength itself, $\kappa$, depends on the filling factor between the FMR mode and the rf $h$-field, and is therefore fixed in our experiment.  Since the resonant frequencies and damping rates of the spin and photon subsystems have already been characterized, the only free parameter is the overall amplitude $A$, which depends on, e.g., the microwave pins and cabelling.  We find $A = 1.22 \times 10^{-3}$ (T/$\mu_0$) for the data in Fig. \ref{fig1} (f).  Excellent agreement between our transient polariton model and the experimental results is evident.

\section{ Time Domain Analysis and Control of the Rabi Oscillations} \label{sec:rabi}

\begin{figure}[!t]
\centering
\includegraphics[width=\linewidth]{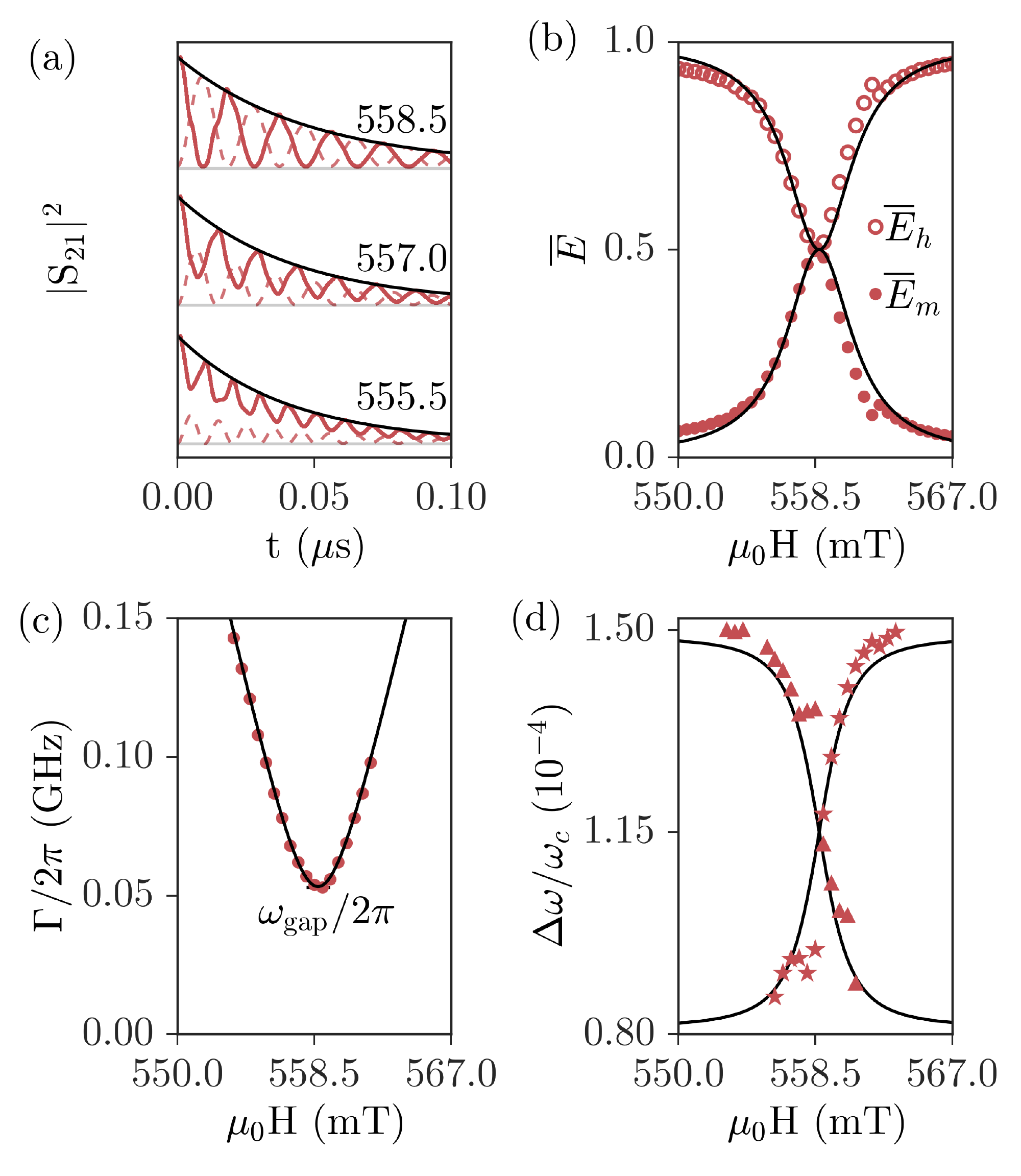}
\caption{Bias field control of the Rabi oscillations.  (a) The raw transmission data ($|\text{S}_{21}|^2 \propto |h|^2$) is displayed as the solid red curve with the $\pi$ phase shifted magnetization $|m|^2$, calculated according to Eq. \eqref{eq:m2}, shown as the dashed red curve.  The black decay curve is the sum, $|m|^2 + |h|^2$, illustrating the total energy decay due to losses.  Both the Rabi oscillation and amplitude are controlled by the bias field which is labelled in mT.  (b) The fractional energy stored in the photon and spin subsystems.  The symbols are fit results while the solid curve is a calculation according to Eqs. \eqref{eq:ehcalc} and \eqref{eq:emcalc}. (c) The generalized Rabi frequency.  The symbols are determined by a fit to Eq. \eqref{eq:h2} and the solid curve is calculated according to Eq. \eqref{eq:rabi}. (d) The decay rates of both polariton modes.  The symbols are fit results and the black curve is the imaginary component calculated from Eq. \eqref{eq:eigen}.}
\label{fig2}
\end{figure}

The solid red curves in Fig. \ref{fig2} (a) show the raw time domain microwave transmission spectra, $|\text{S}_{21}|^2 \propto |h|^2$ at several bias fields.  These spectra are plotted together with the calculated magnetization amplitudes, $|m|^2$, displayed as red dashed curves.  Together these curves give a clear picture of the energy flow within the coupled spin-photon system: at $t = 0$ energy is put into the photon mode.  With time this energy is transferred from the photon to the spin system, oscillating back and forth according to the generalized Rabi-frequency. 
The phase difference between the spin and photon energy flow is $\pi$; that is, when the energy flows out of the photon system it flows into the spin system and vice versa.  Note that this phase difference is fixed for any bias field, unlike the coherence phase between $X_+$ and $X_-$ which will depend on $H$ according to Eq. \eqref{eq:coordtransformc}.  This is because the energy flow compares $|h|^2$ and $|m|^2$, being proportional to the square amplitudes.  The black curve in Fig. \ref{fig2} (a) is the sum $|h|^2 + |m|^2$ which gives a measure of the total energy in the system.  The decay rate of this curve is bias field dependent and related to the damping and fractional energy stored in each spin-photon subsystem.  

For a given bias field the time averaged fractional energy in either the spin or photon subsystem can be calculated according to

\begin{equation}
\overline{E}_m = \frac{\int_0^\infty |m|^2 dt}{\int_0^\infty \left(|h|^2 + |m|^2\right) dt} \label{eq:em}
\end{equation}
with an analogous definition for $\overline{E}_h$.  Therefore using Eqs. \eqref{eq:h2} and \eqref{eq:m2} we find
\begin{equation}
\overline{E}_h = \frac{\left(1-c\right)^2 \Delta \omega_- + c^2\Delta \omega_+}{\left(1-c\right)\Delta \omega_-+c \Delta \omega_+} \approx 1 - 2 c + 2 c^2 \label{eq:ehcalc},
\end{equation}
and
\begin{equation}
\overline{E}_m = \frac{\left(c - c^2\right) \left(\Delta \omega_-+\Delta \omega_+\right)}{\left(1-c\right)\Delta \omega_-+c \Delta \omega_+} \approx 2 \left(c-c^2\right), \label{eq:emcalc}
\end{equation}
where we have used the fact that $\Delta \omega_+ + \Delta \omega_- = \alpha + \beta$ and $\alpha, \beta \ll 1$.  By fitting the transient transmission spectra to Eq. \eqref{eq:h2} we obtain the $c$ parameter, the decay rates $\tau_\pm$ and the generalized Rabi frequency $\Gamma$.  Using the fitted $c$ parameter we can directly determine the time averaged energies, which are plotted as symbols in Fig. \ref{fig2} (b).  The solid curves in Fig. \ref{fig2} (b) are calculated according to Eqs. \eqref{eq:c}, \eqref{eq:ehcalc} and \eqref{eq:emcalc} and agree well with the fit results.  Fig. \ref{fig2} (b) therefore gives a quantitative measure of the fractional energy distribution between spin and photon subsystems and its control via the bias field.  This behaviour is somewhat analogous to that calculated in the phonon-polariton system. \cite{Huang1951a}  However a key difference is that here we are examining the transient response while in Ref. \citenum{Huang1951a} the steady state of the phonon-polariton was examined, with the fractional energy distribution controlled by the wave vector $k$.  The evolution of $\overline{E}_{h, m}$ corresponds to the change in extinction rate observed in the raw data of Fig. \ref{fig2} (a).  At the coincidence point, where the extinction rate is maximized, the energy is equally distributed between the spin and photon subsystems.  Away from $H_c$ the energy transfer is less efficient and most of the polariton energy is stored in the photon subsystem.  This has implications for the total energy decay of the polariton system.  If $\beta > \alpha$ (our case), since more energy is stored in the photon subsystem away from $H_c$, the overall energy loss from the polariton will be greater away from $H_c$.  If $\beta < \alpha$ the situation would be reversed.

The fitting procedure also yields $\Gamma$, and $\tau_\pm = \Delta \omega_\pm^{-1}$.  Fits of the generalized Rabi frequency are plotted as the red symbols in Fig. \ref{fig2} (c) with the black curve a calculation according to Eq. \eqref{eq:rabi}.  Here bias field control of the generalized Rabi frequency is demonstrated by controlling the detuning and explains the frequency changes observed in Fig. \ref{fig2} (a).  At $H_c$, $\delta = 0$ and therefore the pure Rabi frequency is obtained which can be used to extract the coupling strength.  Fig. \ref{fig2} (d) displays the fitted decay rates for each polariton mode as symbols with a calculation of the imaginary component from Eq. \eqref{eq:eigen}.  These decay rates, coupled to the evolution of energy stored in each mode, controls the total decay plotted in Fig. \ref{fig2} (a).

\section{Measuring Mode Composition} \label{sec:comp}

\begin{figure}[!t]
\centering
\includegraphics[width=\linewidth]{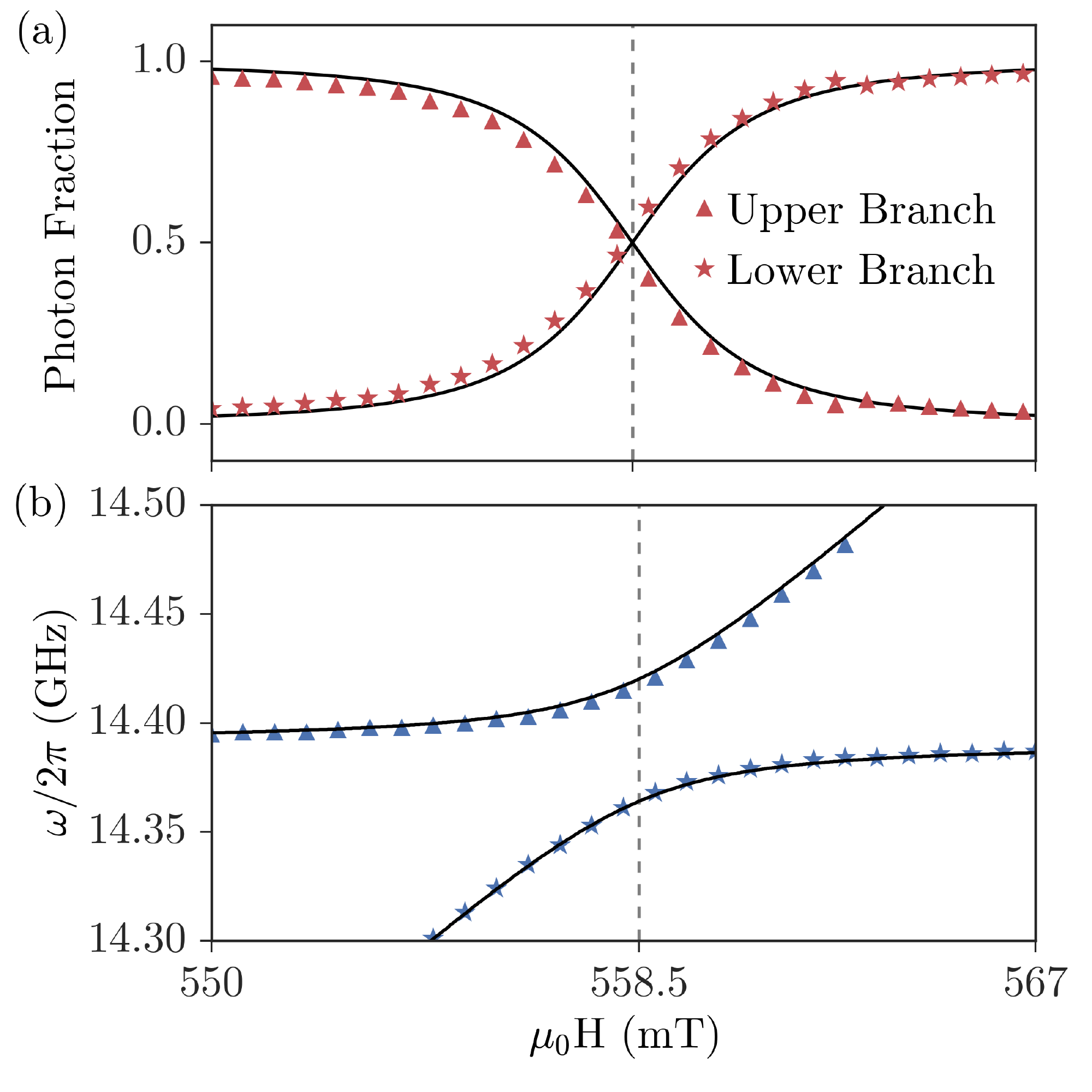}
\caption{(a) Polariton mode composition.  Triangles and stars correspond to the fit results for the amplitudes $C_{11}$, of the upper polariton branch, and $C_{12}$, of the lower polariton branch, respectively.  Solid curves are model calculations according to Eqs. \eqref{eq:coordtransformc} and \eqref{eq:c}.  (b) The polariton dispersions.  At any given bias field, whether the dispersion is photon-like or spin-like will determine the contribution to the photon composition of panel (a).}
\label{fig3}
\end{figure}

By measuring $c$ we have access to the full coordinate transformation which relates the spin/photon and polariton bases.  In particular we can measure the coefficients $C_{11}$ and $C_{12}$ which determine the polariton contributions to the photonic state.  These results are plotted as symbols in Fig. \ref{fig3} (a), normalized to the amplitude $A$.  The triangles and stars indicate the contribution from the upper and lower polariton branches respectively while the solid black curves are calculated according to Eqs. \eqref{eq:coordtransformc} and \eqref{eq:c}.  Here we see the evolution of the photon composition, with the upper branch dominating for $H < H_c$, equal contributions from both polariton branches at $H = H_c$, and the lower branch dominating for $H > H_c$.  
As required the normalized coefficients obey the sum rule $C_{11} + C_{12} = 1$.  This behaviour is analogous to the well known Hopfield coefficients of exciton-polaritons, \cite{KavokinBook} where the composition may be measured e.g. via angle resolved reflectivity spectra. \cite{Liu2014a}

At any given bias field, whether the polariton dispersion is photon-like or spin-like will determine the contribution to the photon composition.  This point can be clarified by comparing the photon composition to the polariton dispersion plotted in Fig. \ref{fig3} (b), where the triangles and stars are fits to the upper and lower branches of the frequency domain transmission spectra, respectively, and the solid curve is calculated using the real part of Eq. \eqref{eq:eigen}.  Both the polariton dispersion and the mode line width, determined by the imaginary part of Eq. \eqref{eq:eigen}, are intrinsic to the coupled system and therefore are identical in the time and frequency domains.  The fit results for the line width have been plotted in Fig. \ref{fig2} (d), but do not play a dominant role in the polariton composition.  This follows from the fact that for level repulsion the damping does not strongly influence the dispersion, although this is no longer true for level attraction. \cite{Harder2018a}  At a given bias field, the polariton branch which is more photon-like will make the largest contribution to the photon composition.  At the crossing point where the polariton dispersion has its maximum deviation from both the spin and photon dispersions, both upper and lower branches contribute equally to the photon composition.   This is the origin of the bias field control of photon composition.   Effectively by controlling the bias field we are controlling the initial ($t = 0$) composition of $h$ and therefore the composition of the photonic state also clarifies the coherent energy exchange found in Fig. \ref{fig2} (b).  
\\[10pt]
\section{Conclusions} \label{sec:conclusions}
Using time domain microwave transmission measurements we have studied the transient response of a coupled spin-photon system.  We find that the Rabi oscillations, which signal coherent spin-photon interactions, can be controlled both in their frequency and amplitude by tuning an external bias field.  This behaviour is accurately modelled by considering the transient response of a two level spin-photon system.  This approach allows us to directly measure the energy exchange between the spin and photon subsystems which comprise the cavity-magnon-polariton.  We find that the efficiency of the energy exchange can be controlled by the bias field, with the energy equally distributed when spin and photon resonance frequencies are coincident.  At this coincidence point we also find that both the upper and lower polariton branches contribute equally to the photon composition.  Therefore we have realized a method both to control and quantify the energy exchange in the cavity-magnon-polariton, allowing us to better understand and use the coherent properties of the polariton, e.g. to minimize total energy losses based on subsystem decay rates and total energy exchange.  Furthermore, the general framework we have developed for time domain analysis of the CMP and a detailed comparison between the transient and CW behaviour is independent of the experimental details, relying instead on an eigenmode analysis, and therefore may benefit future THz time domain studies of CMP phenomena in antiferromagnets.
\section*{Acknowlegements}   

During this work M. H. was partially supported by IODE Canada and an NSERC CGSD Scholarship and P. H. was supported by the UMGF program.  This work was funded by NSERC, CFI, and NSFC (No. 11429401) grants (C.-M.H).  We would like to thank N. L. Wang and J.-Ph. Ansermet for useful discussions.

\bibliography{mainText.bbl}
\end{document}